# An Improved Roe Scheme for All Mach-Number Flows Simultaneously Curing Known Problems

Xue-song Li[*], Xiao-dong Ren, Chun-wei Gu

*Key Laboratory for Thermal Science and Power Engineering of Ministry of Education, Department of Thermal Engineering, Tsinghua University, Beijing 100084, PR China*

**Abstract:** Roe scheme is known for its good performance in moderate-Mach-number flows. However, this scheme and its extended versions suffers from many disastrous problems, such as non-physical behavior, global cut-off, and checkerboard problems, for incompressible flows; and shock instability, expansion shock, and positively non-conservative problems for hypersonic flows. In this paper, non-physical behavior problem, checkerboard problem, and main reason of shock instability problem are due to that the Roe scheme cannot identify multi-dimensional incompressible and compressible flows when normal Mach number on the cell face tends to zero, and then leads to incorrect cross modifications. Positively non-conservative problem is also identified as another important reason for shock instability. Therefore, Mach number and an assistant pressure-density-varying detector are introduced into the Roe scheme to judge compressibility, positivity condition is satisfied by a simple modification with minimal numerical dissipation increases and even with possible decreases in numerical dissipation, the mechanism of the preconditioned Roe scheme is introduced to suppress checkerboard problem, and modified entropy fix and the rotated Riemann solver is combined with complementary advantages as an assistant improvement for better robust. Based on above improvements and previous developments for global cut-off and expansion shock problems, an improvement Roe scheme for all Mach-number flow (Roe-AM) is proposed to simultaneously overcome nearly all well-known drawbacks of the classical Roe scheme. The Roe-AM scheme is simple, easy to implement, computationally low-cost, robust, good extensibility, and free of empirical parameters essentially, with increasing minimal numerical dissipation.

---
* Corresponding author. Tel.: 0086-10-62771209; fax: 0086-10-62771209
  E-mail address: xs-li@mail.tsinghua.edu.cn (X.-S. Li).

**1. Introduction**

Roe scheme [1] is one of the most important shock-capturing schemes and has undergone considerable development because of its good performance for moderate Mach-number flows. However, Roe scheme features several disastrous shortcomings for extremely low- and high-Mach-number flows.

For extremely low-Mach-number flows, that is, incompressible flows, Roe scheme and its low-Mach number versions mainly suffers from all or part of non-physical behavior, global cut-off, and checkerboard problems, which were identified in Refs. [2][3].

In non-physical behavior problem [4], pressure solution of shock-capturing scheme scales with the referenced Mach number at a low-Mach-number speed, that is, $p(x,t) = P_0(t) + M_* p_1(x,t)$, which violates the physical rule on pressure scales with the square of the referenced Mach number, that is, $p(x,t) = P_0(t) + M_*^2 p_2(x,t)$, where $x$ and $t$ denote space and time, respectively.

Traditional all-speed schemes [5][6][7][8][9], which include preconditioning technology and the preconditioned Roe scheme [5][6], focus on curing the non-physical behavior problem besides of accelerating convergence rate. However, the goal cannot be fully achieved because of the global cut-off problem. In this problem, local Mach number in the curing method is replaced with a global reference Mach number, which unfavorably results in limited simulation of mixed flows with low and high Mach numbers. For example, for a flow region where incompressible flows coexist with shock waves, simulation of incompressible regions cannot benefit from the preconditioning technology because of the global cut-off problem. Recently, several improved Roe-type schemes [10][11][12][13][14] were developed to address the global cut-off problem and are summarized by Refs. [2][3].

The checkerboard problem refers to pressure–velocity decoupling in incompressible calculation, which indicates a pressure solution with checkerboard oscillation. The checkerboard problem is an important issue in simulation of incompressible flows and should be suppressed by staggered grids or the momentum interpolation method (MIM) [15][16][17]; otherwise, computation becomes unstable and divergent. Traditional all-speed schemes lack awareness of the checkerboard problem, because they usually unintentionally inherit the curing mechanism, which is similar to MIM [18], from the corresponding shock-capturing scheme. However, several schemes suffer from the checkerboard problem because they cannot inherit the curing mechanism [9][12], or the curing mechanism does not exist for the corresponding shock-capturing scheme [8]. Refs. [2][3] also proposed a general rule for suppressing the checkerboard problem.

For extremely high-Mach-number flows, that is, hypersonic flows, the Roe scheme mainly suffers from some disastrous shortcomings, such as shock instability, expansion shock, and positive non-conservativeness [19].

Shock instability is a well-known defect of supersonic flows and features different performances, such as carbuncle, kinked Mach stem, and odd–even decoupling. To address this issue, various methods have been proposed, and they include combining a dissipative scheme [19], adding an entropy fix [20], increasing basic upwind dissipation [21][22], and considering multi-dimensional characteristics [23][24]. However, these methods constantly introduce large numerical dissipation and unfavorable empirical parameters. Additional, shock instability is due to pressure difference term in mass flux of a scheme [25]. Recently, MIM in the Roe scheme was identified as an important factor in producing shock instability; subsequently, a new method was proposed to address shock instability by decreasing rather than increasing numerical dissipation [26].

Expansion shock is also a Roe scheme defect. This defect violates the entropy condition and is a non-physical solution. Expansion shock usually yields an unacceptable negative pressure and then leads to computation divergence for highly energetic flows. Entropy fix can also be used to overcome this defect. Another solution that is usually adopted provides a slight modification by redefining numerical signal velocities [22][27]. Recently, an improved method was proposed [29], and it is compatible with the shock-instability improvement proposed by Ref. [26].

Positive non-conservativeness is another well-known defect which also results in failure of the Roe scheme on simulating high energetic flows because of the production of negative density and/or temperature. For example, the Roe scheme cannot converge in near-vacuum flow with low density because of positive non-conservativeness [27]. Therefore, robustness of the Roe scheme is weakened. Traditionally, no independent analysis and method satisfy the positive condition of the Roe scheme, and preserving positive conservativeness is usually an additional function of curing shock instability. Typically, entropy fix can simultaneously resolve this problem to a significant extent.

In summary, significant efforts have been exerted to address low-Mach-number and high-Mach-number problems of the Roe scheme. However, no developed scheme can simultaneously address these problems. Therefore, significant interest on and demand for developing such a scheme exist. Such scheme can suppress all known problems without introducing unfavorable characteristics, such as large numerical dissipation and empirical parameters. This goal motivates us to propose a Roe-type scheme for all Mach-number flows (Roe-AM) which can simultaneously cure known problems.

The rest of this paper is organized as follows. Chapter 2 provides the governing equations and reviews the Roe scheme. Chapter 3 reviews the general rules for extending the Roe scheme to incompressible flows and proposes a simple method for obtaining an all-Mach-number scheme and its asymptotic analysis. Chapter 4 analyzes mechanism of problems of the Roe scheme, especially for the positivity condition. Chapter 5 proposes further improvement for robust by combing modified entropy fix and the rotated Riemann solver with limited increased dissipation. Chapter 6 proposes the Roe-AM scheme and some remarks. Chapter 7 provides classical numerical cases to validate the Roe-AM scheme and demonstrates that positivity condition is the sole cause of shock instability problem. Chapter 8 concludes this paper.

## 2. Governing Equations and the Roe Scheme

### 2.1 Governing Equations

The governing three-dimensional Navier–Stokes equation can be written as follows:

$$\frac{\partial Q}{\partial t} + \frac{\partial F}{\partial x} + \frac{\partial G}{\partial y} + \frac{\partial H}{\partial z} = 0, \qquad (1)$$

where $Q = \begin{bmatrix} \rho \\ \rho u \\ \rho v \\ \rho w \\ \rho E \end{bmatrix}$ is vector of conservation variables; $F = \begin{bmatrix} \rho u \\ \rho u^2 + p \\ \rho uv \\ \rho uw \\ \rho uH \end{bmatrix}$, $G = \begin{bmatrix} \rho v \\ \rho uv \\ \rho v^2 + p \\ \rho vw \\ \rho vH \end{bmatrix}$, and $H = \begin{bmatrix} \rho w \\ \rho uw \\ \rho vw \\ \rho w^2 + p \\ \rho wH \end{bmatrix}$ are vectors of Euler fluxes; $\rho$ is fluid density; $p$ is pressure; $E$ is total energy; $H$ is total enthalpy; $u$, $v$, and $w$ respectively correspond to velocity components in Cartesian coordinates $(x, y, z)$.

### 2.2 Classical Roe Scheme

The classical Roe scheme can be expressed as the sum of a central term and a numerical dissipation term, as shown in the following general form:

$$\tilde{F} = \tilde{F}_c + \tilde{F}_d, \qquad (2)$$

where $\tilde{F}$ is the numerical flux, $\tilde{F}_c$ represents the central term, and $\tilde{F}_d$ is the numerical dissipation term. For the cell face of finite volume method, the equations used are as

follows:

$$\tilde{F}_{c,1/2} = \frac{1}{2}(\bar{F}_L + \bar{F}_R),\qquad(3)$$

$$\bar{F} = U\begin{bmatrix}\rho\\\rho u\\\rho v\\\rho w\\\rho H\end{bmatrix} + p\begin{bmatrix}0\\n_x\\n_y\\n_z\\0\end{bmatrix},\qquad(4)$$

where $n_x$, $n_y$, and $n_z$ refer to the components of face-normal vector, and $U = n_x u + n_y v + n_z w$ is normal velocity on the cell face.

The numerical dissipation term, $\tilde{F}_d$, is usually expressed in vector form as follows:

$$\tilde{F}_{d,\frac{1}{2}}^{Roe} = -\frac{1}{2}R_{\frac{1}{2}}^{Roe}\Lambda_{\frac{1}{2}}^{Roe}\left(R_{\frac{1}{2}}^{Roe}\right)^{-1}\Delta Q,\qquad(5)$$

where $\Delta Q = Q_R - Q_L$, $R^{Roe}$ is the right eigenvector matrix of the governing equations, and $\Lambda^{Roe}$ is the diagonal matrix formed with the relevant eigenvalues:

$$\lambda_1 = \lambda_2 = \lambda_3 = |U|,\ \lambda_4 = |U-c|,\ \lambda_5 = |U+c|,\qquad(6)$$

where $c$ is the sound speed. For simplicity, in the case of no confusion subscript "$_{1/2}$" will be omitted.

## 2.3 Scale Framework of the Roe-type Scheme

Eq. (5) can also be expressed as an equivalent scale algorithm [3][6] as follows:

$$\tilde{F}_d = -\frac{1}{2}\left\{\xi\begin{bmatrix}\Delta\rho\\\Delta(\rho u)\\\Delta(\rho v)\\\Delta(\rho w)\\\Delta(\rho E)\end{bmatrix} + (\delta p_u + \delta p_p)\begin{bmatrix}0\\n_x\\n_y\\n_z\\U\end{bmatrix} + (\delta U_u + \delta U_p)\begin{bmatrix}\rho\\\rho u\\\rho v\\\rho w\\\rho H\end{bmatrix}\right\},\qquad(7)$$

where $\xi$ is the basic upwind dissipation, $\delta p_p$ is the pressure-difference-driven modification for the cell face pressure, $\delta p_u$ denotes velocity-difference-driven modifications for the cell face pressure, $\delta U_u$ indicates the velocity-difference-driven modification for the cell face velocity, and $\delta U_p$ specifies the pressure-difference-driven modification for cell face velocity.

Eq. (7) can be regarded as a uniform framework for the shock-capturing scheme [3], which is simple, computationally inexpensive, and easy to analyze and improve. Therefore, Eq. (7) serves as the basis of this paper. Based on the equation, the classical Roe scheme is expressed as follows:

$$\xi = \lambda_1, \tag{8}$$

$$\delta p_u = \left(\frac{\lambda_5 + \lambda_4}{2} - \lambda_1\right)\rho \Delta U, \tag{9}$$

$$\delta p_p = \frac{\lambda_5 - \lambda_4}{2}\frac{\Delta p}{c}, \tag{10}$$

$$\delta U_u = \frac{\lambda_5 - \lambda_4}{2}\frac{\Delta U}{c}, \tag{11}$$

$$\delta U_p = \left(\frac{\lambda_5 + \lambda_4}{2} - \lambda_1\right)\frac{\Delta p}{\rho c^2}. \tag{12}$$

Based on Eq. (6), Eqs. (8)–(12) can be further simplified as follows:

$$\xi = |U|, \tag{13}$$

$$\delta p_u = \max\left(0, c - |U|\right)\rho \Delta U, \tag{14}$$

$$\delta p_p = \text{sign}(U)\min\left(|U|, c\right)\frac{\Delta p}{c}, \tag{15}$$

$$\delta U_u = \text{sign}(U)\min\left(|U|, c\right)\frac{\Delta U}{c}, \tag{16}$$

$$\delta U_p = \max\left(0, c - |U|\right)\frac{\Delta p}{\rho c^2}. \tag{17}$$

## 3. Simple Method for Constructing All-Mach Roe Scheme

## 3.1 General Rules for Extending the Roe scheme to Incompressible Flows

For extending the Roe scheme to incompressible flows, the following rules should be observed when Mach number approaches zero, that is, $M \to 0$ [2][3]:

(1) For overcoming the non-physical behavior problem, $\delta p_u \leq O(\rho u \Delta u)$;

(2) For overcoming the global cut-off problem, global cut-off in the numerator can be directly cancelled although it can be retained in the denominator;

(3) For overcoming the checkerboard problem, $\delta U_p = O(c^{-1}\rho^{-1}\Delta p \sim u^{-1}\rho^{-1}\Delta p)$.

As demonstrated in Ref. [2], the first and second rules are no problem, but the third rule indicates an unenviable choice. When $\delta U_p = O(c^{-1}\rho^{-1}\Delta p)$, the typical of which is the Roe scheme as shown in Eq. (17), it can enforce a divergence-free constraint of the leading-order velocity but suffers from weak checkerboard. When $\delta U_p = O(u^{-1}\rho^{-1}\Delta p)$, the typical of which is the preconditioned Roe scheme [2][4][5][6], on the contrary, the checkerboard problem can be solved well but divergence constraint is unsatisfied. Ref. [28] also discusses this problem and suggests the latter that does not enforce the discrete divergence free constraint for better suppressing the checkerboard modes.

Therefore, in this paper, for satisfying the third rule with $\delta U_p = O(u^{-1}\rho^{-1}\Delta p)$, the $\delta U_p$ term adopts the term of the preconditioned Roe scheme as follows:

$$\delta U_p = \left[ \max\left(0, \tilde{c} - |U|\right) + (1-\theta)\left(|U| - \frac{U}{2\tilde{c}}\text{sign}(\tilde{U})\min\left(|\tilde{U}|, \tilde{c}\right)\right) \right] \frac{\Delta p}{\rho \theta c^2}, \qquad (18)$$

$$\tilde{c} = \frac{1}{2}\sqrt{4c^2\theta + (1-\theta)^2 U^2}, \qquad (19)$$

$$\tilde{U} = \frac{1}{2}(1+\theta)U, \qquad (20)$$

$$\theta = \min\left[\max\left(M_{ref}^2, M^2\right), 1\right]. \qquad (21)$$

where $M_{ref}$ is a global reference Mach number such as main flow or maximum Mach number of flow filed. Eq. (18) becomes Eq. (17) when $M_{ref} \geq 1$ or $M \geq 1$.

Additional, the non-physical behavior problem can be avoided by adding a simple improvement for $\delta p_u = O(u \Delta u)$ as follows:

$$\delta p_u = f(M) \max(0, c - |U|) \rho \Delta U, \qquad (22)$$

where $f(M)$ serves as a function of the Mach number as follows:

$$f(M) = \min(O(M^n), 1). \qquad (23)$$

When $n \geq 1$, (24)

the first rule can be satisfied. An expression is proposed as follows [2][12][29][30]:

$$f(M) = \min\left(M \frac{\sqrt{4 + (1-M^2)^2}}{1+M^2}, 1\right), \qquad (25)$$

because it features a low value when $M \to 0$ and a smooth transition when $M \to 1$.

The second rule can be satisfied automatically except the $\delta U_p$ term, which is a necessary slight cost for controlling the checkerboard problem well.

Therefore, by replacing the $\delta p_u$ term Eq. (14) and the $\delta U_p$ term Eq. (17) with Eq. (22) and Eq. (18), respectively, the improved Roe scheme can be extended to incompressible flows.

**3.2 Asymptotic Analysis**

In order to better understand mechanism of above discussions in section 3.1, an asymptotic analysis, which is a powerful tool to analyze low Mach number behaviour [4], is given as follows.

For asymptotic analysis, all non-dimensional variables are asymptotically expanded into powers of the reference Mach number $M_*$:

$$\tilde{\phi} = \tilde{\phi}^0 + M_*\tilde{\phi}^1 + M_*^2\tilde{\phi}^2 + M_*^3\tilde{\phi}^3 + \cdots, \tag{26}$$

where $\phi$ represents one of the fluid variables, $\rho$, $u$, $v$, $E$, or $p$.

On substituting these non-dimensional variables into governing equations, the dimensionless discrete equations can be obtained and terms of equal power of $M_*$ can be collected. From the asymptotically expanded momentum equation, the following terms should be obtained for the physical behaviour of low-Mach-number flows:

$$p_{i-1,j}^0 - p_{i+1,j}^0 = 0, \tag{27}$$

$$p_{i,j-1}^0 - p_{i,j+1}^0 = 0, \tag{28}$$

$$p_{i-1,j}^1 - p_{i+1,j}^1 = 0, \tag{29}$$

$$p_{i,j-1}^1 - p_{i,j+1}^1 = 0. \tag{30}$$

Eqs. (27) – (30) result in $p_i^0 = cte \; \forall i$ and $p_i^1 = cte \; \forall i$, which means that pressure fluctuation is order of $p_2$ and scales with the square of the Mach number as follows:

$$p(x,t) = P_0(t) + M_*^2 p_2(x,t). \tag{31}$$

For the classical Roe scheme, it is proved that Eqs. (29)-(30) cannot be satisfied and then pressure fluctuation is order of $p_1$ as follows:

$$p(x,t) = P_0(t) + M_* p_1(x,t). \tag{32}$$

It is why the non-physical behavior problem occurs.

When the first rule in section 3.1 is adopted such as using Eqs. (22) and (25), the non-physical behavior problem can be solved by satisfying Eqs. (27) – (30) and then recovering the Eq. (31).

However, except a constant, a chess-like four-field solution of $p_0$ and $p_1$ also satisfies Eqs. (27) – (30). It is why the checkerboard problem occurs and the third rule in section 3.1 should be considered.

When $\delta U_p = \mathrm{O}(c^{-1}\rho^{-1}\Delta p)$ such as Eq. (17) of the Roe scheme, additional constraints are produced as follows:

$$\sum_{l \in v(i)} \frac{\Delta_{il} p^0}{c_{il}^0} = 0, \tag{33}$$

$$u_{i+1,j}^0 - u_{i-1,j}^0 + v_{i,j+1}^0 - v_{i,j-1}^0 = -\sum_{l \in v(i)} \frac{\Delta_{il} p^1}{c_{il}^0} = 0. \tag{34}$$

Eq. (33) provides a constraint suppressing the four-field solution of $p_0$. However, the solution of $p_1$ is lack of similar mechanism. Therefore, the pressure solution suffers from weak checkerboard problem due to $p_1$. Excluding the possibility of the four-field solution, Eq. (34) is equal to zero and then enforces a divergence-free constraint of the leading-order velocity as discussed in section 3.1. Considering the checkerboard, however, Eq. (34) is unsatisfied in fact.

When $\delta U_p = \mathrm{O}(u^{-1}\rho^{-1}\Delta p)$ such as Eq. (18) of the preconditioned Roe scheme, additional constraints are produced as follows from the asymptotically expanded continuity or energy equation:

$$\sum_{l \in v(i)} \frac{\Delta_{il} p^0}{\tilde{c}_{il}^0} = 0, \tag{35}$$

$$\sum_{l \in v(i)} \frac{\Delta_{il} p^1}{\tilde{c}_{il}^0} = 0, \tag{36}$$

$$u_{i+1,j}^0 - u_{i-1,j}^0 + v_{i,j+1}^0 - v_{i,j-1}^0 = -\sum_{l \in v(i)} \frac{\Delta_{il} p^2}{\tilde{c}_{il}^0} \neq 0. \tag{37}$$

Eqs. (35)-(36) provide a constraint suppressing the four-field solution of $p_0$ and $p_1$. Therefore, the checkerboard problem is cured. However, Eq. (35) means that the divergence-free constraint of the leading-order velocity is unsatisfied.

Therefore, in this paper Eq. (18) is adopted to cure the checkerboard problem, because it has ability to avoid all checkerboard modes and it maybe not worse than Eq. (17) for enforcing divergence-free of velocity.

### 3.3 A Simple Method for Constructing the All-Mach Roe Scheme

As discussed in sections 3.1 and 3.2, the Roe scheme is well extended to incompressible flows by replacing Eqs. (14) and (17) with Eqs. (22) and (18), respectively. For an all-Mach-number scheme, however, it is still lacks a good shock-stability fix.

According to [11][12], a simple method for constructing the all-Mach Roe scheme can be proposed, and it combines a version of the Roe scheme for low-Mach-number flows and another version for supersonic flows through a function of the local Mach number $f(M)$ as follows:

$$\tilde{\boldsymbol{F}}^{AM-Roe} = \left[1 - f(M)\right]\tilde{\boldsymbol{F}}^{LM-Roe} + f(M)\tilde{\boldsymbol{F}}^{S-Roe}, \qquad (38)$$

where $\tilde{\boldsymbol{F}}^{LM-Roe}$ is any Roe-type scheme with a low-Mach-number fix, such as Eqs. (18) and (22); and $\tilde{\boldsymbol{F}}^{S-Roe}$ is any Roe-type scheme with shock-stability fix for supersonic flows, such as the methods discussed in Refs. [19]–[29]; the requirement for $f(M)$ is the same as in Eqs. (23) and (24), such that in Eq. (25). $\tilde{\boldsymbol{F}}^{AM-Roe}$ becomes $\tilde{\boldsymbol{F}}^{LM-Roe}$ when $M \to 0$ and $\tilde{\boldsymbol{F}}^{S-Roe}$ when $M \geq 1$.

Eq. (22) is a simple method for obtaining the scheme for all-Mach-number flows on the basis of existing methods. However, this equation cannot discover the mechanism of shock problems and inherits the defects of existing methods; such defects include increasing numerical dissipation to suppress shock instability.

Therefore, construction of an all-Mach scheme is still worth studying. The following section proposes an ideal all-Mach Roe scheme by analyzing and obtaining new insights into the problem mechanisms and making subtle changes to the classical Roe scheme.

## 4. Positivity Condition of the Roe Scheme

Traditionally, positive non-conservativeness is regarded as the cause for computational divergence for near-vacuum flow. In this paper, positive non-conservativeness is also identified as the cause of shock instability in the following sections.

### 4.1 Positivity Condition

In the positivity condition, scalar quantities should preserve positivity [27], especially when velocity approaches zero. In the following sections, only density is discussed in the positivity condition because it can represent other scalar quantities and can be easily analyzed from the continuity equation. Therefore, the positivity condition can be expressed as the numerical flux form of continuity equation as follows:

$$\tilde{F}^{continuity} = \rho^* u = a_1 u_1 \rho_L + a_2 u_2 \rho_R, \qquad (39)$$

where the following expressions should be satisfied to ensure positivity:

$$a_1 \geq 0 \text{ and } a_2 \geq 0. \qquad (40)$$

For simplicity, in the following sections, only the condition $U \geq 0$ is considered, because the same conclusion can be obtained when $U \leq 0$, and detailed discussion is omitted.

### 4.2 Positivity Condition when $U \geq c$

When $U \geq c$, numerical fluxes, $\tilde{F}$, of the Roe scheme can be simplified as follows:

$$\tilde{F}_{0.5} = \tilde{F}_c + \tilde{F}_d = p_L \begin{bmatrix} 0 \\ n_x \\ n_y \\ n_z \\ 0 \end{bmatrix}_{0.5} + \frac{U_R - U_{0.5}}{2} \begin{bmatrix} \rho \\ \rho u \\ \rho v \\ \rho w \\ \rho H \end{bmatrix}_R + \frac{U_L + U_{0.5}}{2} \begin{bmatrix} \rho \\ \rho u \\ \rho v \\ \rho w \\ \rho H \end{bmatrix}_L - \frac{U_R - U_L}{2} \begin{bmatrix} \rho \\ \rho u \\ \rho v \\ \rho w \\ \rho H \end{bmatrix}_{0.5} \quad (41)$$

If the following expressions can be satisfied:

$$U_{0.5} = \frac{1}{2}(U_R + U_L), \quad (42)$$

$$\begin{bmatrix} \rho \\ \rho u \\ \rho v \\ \rho w \\ \rho H \end{bmatrix}_{0.5} = \frac{1}{2} \begin{bmatrix} \rho \\ \rho u \\ \rho v \\ \rho w \\ \rho H \end{bmatrix}_L + \frac{1}{2} \begin{bmatrix} \rho \\ \rho u \\ \rho v \\ \rho w \\ \rho H \end{bmatrix}_R, \quad (43)$$

full upwind scheme can be obtained as follows:

$$\tilde{F}_{1/2} = \tilde{F}_L. \quad (44)$$

Then, positivity condition and full upwind can be satisfied; satisfaction of full upwind is required for supersonic flows because downstream information cannot be passed upstream.

However, Eqs. (42) and (43) cannot be satisfied simultaneously. Moreover, Roe average is usually adopted to evaluate averaged values, which visibly deviate from the simple average in Eqs. (42) and (43). Therefore, an unexpected conclusion can be obtained: the classical Roe scheme cannot satisfy full upwind and positivity requirements.

Although Eqs. (42) and (43) cannot be satisfied, this difficulty can be easily overcome by directly introducing Eqs. (42) and (43) into the numerical dissipation of the classical Roe scheme in Eq. (7). Therefore, the improved expression, which satisfies the positivity condition and full upwind requirement, is given as follows:

$$\tilde{F}_{d,0.5} = -\frac{1}{2}\left\{ \xi \begin{bmatrix} \Delta\rho \\ \Delta(\rho u) \\ \Delta(\rho v) \\ \Delta(\rho w) \\ \Delta(\rho H) \end{bmatrix} + \left(\delta p_p + \delta p_u\right)\begin{bmatrix} 0 \\ n_x \\ n_y \\ n_z \\ 0 \end{bmatrix} + \delta U_p \begin{bmatrix} \rho \\ \rho u \\ \rho v \\ \rho w \\ \rho H \end{bmatrix}_{0.5} \right\} - \frac{1}{2}\delta U_{u,R}\begin{bmatrix} \rho \\ \rho u \\ \rho v \\ \rho w \\ \rho H \end{bmatrix}_R - \frac{1}{2}\delta U_{u,L}\begin{bmatrix} \rho \\ \rho u \\ \rho v \\ \rho w \\ \rho H \end{bmatrix}_L$$

, (45)

where definitions of $\delta p_u$, $\delta p_p$, and $\delta U_p$ remain unchanged, except for $\xi$, $\delta U_{u,R}$, and $\delta U_{u,L}$, which are as follows:

$$\xi = \frac{|U_R + U_L|}{2}, \quad (46)$$

$$\delta U_{u,R} = \frac{1}{2}\text{sign}(U)\min(|U|, c)\frac{\Delta U}{c}, \quad (47)$$

$$\delta U_{u,L} = \frac{1}{2}\text{sign}(U)\min(|U|, c)\frac{\Delta U}{c}. \quad (48)$$

In this paper, numerical cases do not show difference between Eq. (7) and Eq. (45) subdividing $\delta U_u$ into $\delta U_{u,R}$ and $\delta U_{u,L}$. Based on theoretical consideration of full upwind, Eq. (45) is adopted in the paper.

In addition, total enthalpy conservation should also be preserved. This requirement can be satisfied by introducing the following condition to the numerical dissipation constraint of continuity and energy equations [31][32]:

$$\tilde{F}_d^{energy} = \tilde{F}_d^{continuity} \times H. \tag{49}$$

On the basis of Eq. (49), the energy equation in Eq. (45) is modified by replacing $\Delta(\rho E)$ in the first term of the right side of Eq. (7) with $\Delta(\rho H)$ and $U$ in the second term with 0, according to Refs. [21][22].

### 4.3 Positivity Condition when $U \to 0$

The positivity has key effect under the condition of $U \to 0$, especially when $M > 1$. In the following sections it is analyzed.

When $U \to 0$ and $U \geq 0$, numerical flux of continuity equation of the classical Roe scheme is expressed as follows:

$$\tilde{F}^{continuity} = \tilde{F}_{\rho,u} + \tilde{F}_{\rho,p}, \tag{50}$$

where $\tilde{F}_{\rho,u}$ is defined as the velocity-positivity term, and $\tilde{F}_{\rho,p}$ is the pressure-positivity term:

$$\tilde{F}_{\rho,u} = \frac{U_R - U_L}{4}\rho_R + \frac{U_R + 3U_L}{4}\rho_L, \tag{51}$$

$$\tilde{F}_{\rho,p} = -\left[\max\left(0, \tilde{c} - |U|\right) + (1-\theta)\left(|U| - \frac{\tilde{U}}{2\tilde{c}}U\right)\right]\frac{\Delta p}{2\theta c^2}, \tag{52}$$

where the pressure positivity term, $\tilde{F}_{\rho,p}$, originates from the term $\delta U_p$ in Eq. (18), and the velocity-positivity term $\tilde{F}_{\rho,u}$ originates from the term $\xi$ in Eq. (46) and the central term $\tilde{F}_c$ in Eq. (3).

#### 4.3.1 Analysis of Pressure-Positivity Term

When $U \to 0$, two possible conditions exist: $M \to 0$ and $M > 1$. The two conditions should be analyzed separately.

When $M > 1$, Eq. (52) becomes

$$\tilde{F}_{\rho,p} = -\frac{\Delta p}{2c}. \tag{53}$$

Ref. [25] proposes a conjecture that a scheme is a shock-stable scheme when its pressure difference term in mass flux is equal to zero, i.e., $\tilde{F}_{\rho,p} = 0$. Further, in Ref. [26], the term $\delta U_p$ is regarded as an important cause of shock instability from the viewpoint of momentum interpolation mechanism, and should be equal to zero when $M > 1$. However, a theoretical analysis was not provided in Ref. [26].

In this paper, the said cause is explained from the positivity condition perspective as follows.

Given that

$$p = \gamma^{-1} \rho c^2, \tag{54}$$

where $\gamma$ is adiabatic index,

$$\Delta p = \gamma^{-1}\left(2\rho c \Delta c + c^2 \Delta \rho\right). \tag{55}$$

Substituting Eq. (55) into Eq. (53), we obtain the following:

$$\tilde{F}_{\rho,p} = -\frac{1}{2\gamma}\left(2\rho \Delta c + c \Delta \rho\right) \approx \frac{-c_{1/2} - \Delta c}{2\gamma}\rho_R + \frac{c_{1/2} - \Delta c}{2\gamma}\rho_L. \tag{56}$$

The pressure-difference term, $\Delta p$, introduces negativity terms, especially for $\rho_R$ with a large coefficient of order of $-c$. Therefore, pressure-positivity term $\tilde{F}_{\rho,p}$ inevitably destroys the positivity condition. This situation is the reason why term $\delta U_p$ should be zero under the condition of compressible flows, i.e., $M > 1$ or $M \to 1$.

When $M \to 0$, the situation becomes different. Eq. (52) becomes

$$\tilde{F}_{\rho,p} \approx -\frac{\Delta p}{2\tilde{c}} \approx -\frac{\Delta p}{2u_{ref}}. \tag{57}$$

Given that flows become incompressible, because pressure fluctuation scales with the square of the Mach number as shown in Eq. (31), the following equation can be obtained according to the Bernoulli equation:

$$\Delta p = -\rho u \Delta u. \tag{58}$$

Therefore, by substituting Eq. (58) into Eq. (57), the following is obtained:

$$\tilde{F}_{\rho,p} \approx \frac{u\Delta u}{4u_{ref}}\rho_R + \frac{u\Delta u}{4u_{ref}}\rho_L. \tag{59}$$

Eq. (59) introduces some negative influence for the positivity condition. On the one hand, however, Eq. (59) exerts little effect on the positivity property because $u \leq u_{ref}$ in general. On the other hand, the positivity condition maybe not important for incompressible flows as shown in numerical cases in this paper.

From another perspective, $\delta U_p$ is necessary when $M \to 0$ for suppressing the checkerboard problem, as demonstrated in section 3.2. These conditions mean that $\delta U_p$ can be and should be kept.

Therefore, for compressible flows, $\delta U_p$ should be equal to zero, and for incompressible flows, $\delta U_p$ of the preconditioned Roe scheme, Eq. (18), should be retained. Therefore, criteria are needed to judge whether the flow is compressible or incompressible. Mach number is a good criterion, but it is not enough for certain conditions because of the possibility of low-Mach-number compressible flows, such as the low-Mach-number cells in numerical shock. Because the pressure variation is small as shown in Eq. (31), a detector for pressure and density change maybe a good assistant criterion. A detector discussed in Ref. [22] can satisfy the requirement, and is given as follows with some improvement:

$$s_1 = f^8(a), \tag{60}$$

$$a_{i+1/2,j,k} = \min\begin{pmatrix} P_{i+1/2,j,k}, P_{i+1,j-1/2,k}, P_{i+1,j+1/2,k}, P_{i,j-1/2,k}, P_{i,j+1/2,k}, \\ P_{i+1,j,k-1/2}, P_{i+1,j,k+1/2}, P_{i,j,k-1/2}, P_{i,j,k+1/2} \end{pmatrix}, \tag{61}$$

$$P_{i+1/2,j,k} = \min\left(\frac{p_{i,j,k}}{p_{i+1,j,k}}, \frac{p_{i+1,j,k}}{p_{i,j,k}}, \frac{\rho_{i,j,k}}{\rho_{i+1,j,k}}, \frac{\rho_{i+1,j,k}}{\rho_{i,j,k}}\right), \tag{62}$$

where function $f$ is defined by Eq. (25), and function $a$ is a pressure-density-varying detector by assessing current $i+1/2$ cell face and its eight neighbor faces. In order to ensure that the detector $s_1$ remains nearly zero for large pressure or density change, and has a smooth transition near small change, the function $f^8$ [26] is adopted as shown in Fig. 1.

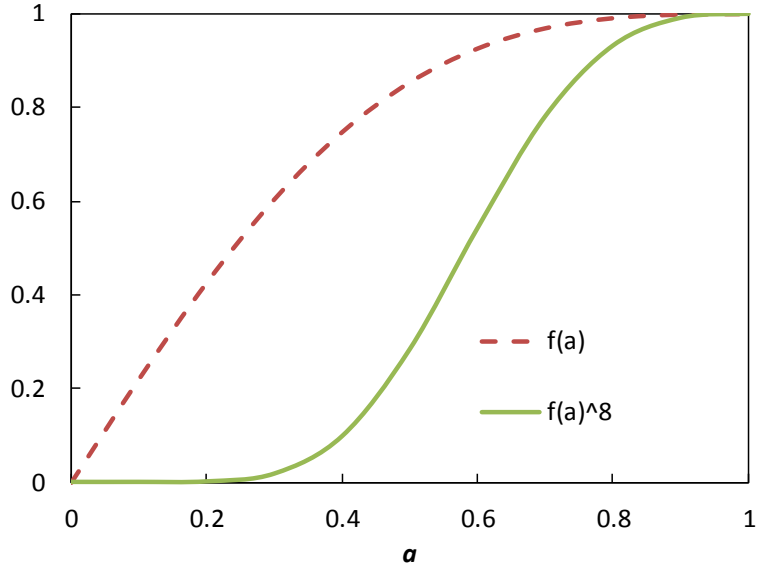

Fig. 1 Functions of $f$ and $f^8$

Therefore, the corresponding improvement for $\delta U_p$ based on Ref. [26] is expressed as follows:

$$\delta U_p = s_1\left[1 - f^8(M)\right]\left[\max\left(0, \tilde{c} - |U|\right) + (1-\theta)\left(|U| - \frac{U}{2\tilde{c}}\mathrm{sign}(\tilde{U})\min(|\tilde{U}|, \tilde{c})\right)\right]\frac{\Delta p}{\rho \theta c^2}$$

,                                                                                     (63)

where coefficient $s_1\left[1-f^8(M)\right]$ makes $\delta U_p = 0$ when flow is compressible, i.e., $M \to 1$ or $M > 1$ or pressure or density has large variation, and $\delta U_p$ is retained only when flow is incompressible, i.e., $M \to 0$ and pressure- and density-variation are small.

### 4.3.2 Analysis of Velocity-Positivity Term

Aside from the pressure-positivity term, the velocity-positivity term (Eq. (51)) also violates the positivity condition, because the first term at the right side includes a negative $U_L$, which makes Eqs. (39) and (40) unsatisfied.

To address positivity non-conservation of the velocity-positivity term, an improvement can be proposed as follows:

$$\xi = \left|\frac{U_R + U_L}{2}\right| + \frac{\Delta U}{2}. \tag{64}$$

Adopting Eq. (64), the following numerical fluxes of continuity equation are provided when $U \to 0$ and $U > 0$ and $M > 1$:

$$\tilde{F}^{\text{continuity}} = \frac{U_R + U_L}{2}\rho_L, \tag{65}$$

and when $U \to 0$ and $U < 0$ and $M > 1$:

$$\tilde{F}^{\text{continuity}} = \frac{U_R + U_L}{2}\rho_R. \tag{66}$$

Therefore, Eqs. (65) and (66) satisfy the positivity condition of Eqs. (39) and (40). This condition implies that positivity condition can be satisfied with a slight adjustment. Compared with Eq. (46), numerical dissipation only increases slightly when $\Delta U > 0$ and even decreases when $\Delta U < 0$.

In order to prevent a negative value of dissipation for a special condition of $U_R + U_L = 0$ and $\Delta U < 0$, and consider possible further improvement, Eq. (64) can be further improved as follows:

$$\xi = \max\left(\left|\frac{U_R + U_L}{2}\right| + \frac{\Delta U}{2}, 0\right), \tag{67}$$

Positivity condition of the velocity-positivity term can also be satisfied by adopting Eq. (67). Considering recovering $\xi$ to Eq. (46) when $U \geq c$:

$$\xi = \max\left\{\left|\frac{U_R + U_L}{2}\right| + \left[1 - f^8(\bar{M})\right]\frac{\Delta U}{2}, 0\right\}, \tag{68}$$

$$\bar{M} = \frac{|U|}{c}, \tag{69}$$

where function $f$ is defined by Eq. (25), and $\bar{M}$ is normal Mach number on the cell face.

### 4.4 Expansion Shock

As discussed in Ref. [26], the term $\delta U_p$ of the classical Roe scheme functions in smoothing physical or non-physical shock. Therefore, improvement of Eq. (63) by removing term $\delta U_p$ can prevent destruction of physical compression shock. However, Eq. (63) retains and deteriorates the non-physical expansion shock and makes the traditional curing method for the expansion shock invalid. Therefore, an improved method is proposed in Ref. [29]; the method is compatible with Eq. (63) and is effective for suppressing expansion shock. The improvement replaces $|U|$ in terms $\delta p_p$, $\delta p_u$, $\delta U_u$, and $\delta U_p$ with $|U|'$ defined as follows:

$$|U|' = |U| - \frac{\text{sign}(U+c)\max(0, U_R - U_L) - \text{sign}(U-c)\max(0, U_R - U_L)}{4}. \tag{70}$$

If $U$ is defined as follows:

$$U = \frac{U_R + U_L}{2}, \quad (71)$$

Eq. (70) can also be expressed as follows:

$$|U|' = \begin{cases} \min(|U_L|, |U_R|) & |U| < c \text{ and } U_R > U_L \\ |U| & \text{otherwise} \end{cases}. \quad (72)$$

Thus, for subsonic expansion flows, the value of $|U|'$ decreases within a reasonable range. Then, $\delta p_u$ and $\delta U_p$ increase, and $\delta p_p$ and $\delta U_u$ decrease synchronously.

To minimize the impact of extra modification, Eq. (70) can also be further improved as follows, because expansion shock occurs when $|U| \to c$:

$$|U|' = |U| - f^8(\bar{M}) \frac{\text{sign}(U+c)\max(0, U_R - U_L) - \text{sign}(U-c)\max(0, U_R - U_L)}{4}. \quad (73)$$

## 5. Further Improvement for Robust by Increasing Limited Dissipation

In section 4.3.2, the basic upwind dissipation term $\xi$ can be slightly adjusted to ensure the positivity condition. This adjustment slightly increases or decreases numerical dissipation. Remarkably increasing the term $\xi$ is another method, which is simple, rough and covers inner mechanism of problems, but is effective and robust [20]-[24], especially for cell faces nearly parallel to flow which is usually regarded as less dissipation. Therefore, in order to make scheme more robust with a limited cost of increasing dissipation, a further improvement for $\xi$ is proposed by combing popular entropy fix and rotated Riemann solvers.

Considering a modification similar to the entropy fix, the Eq. (67) can be improved as follows:

$$\xi = \max\left(\left|\frac{U_R + U_L}{2}\right| + \frac{\Delta U}{2}, \varepsilon_1 \left[1 - f^8(\bar{M})\right] f^8(M) f_{ef}\right), \tag{74}$$

$$f_{ef} = \varepsilon_2 c, \tag{75}$$

where the coefficient $\varepsilon_1$ is equal to 0 or 1 to decide whether the modification is introduced. The function $\left[1 - f^8(\bar{M})\right] f^8(M)$ indicates that the modification is activated only when $\bar{M} \to 0$ and $M \to 1$ or $M \geq 1$, i.e., cell faces nearly parallel to high Mach-number flow. The modification function $f_{ef}$ is simpler than classical entropy fix, and empirical parameter $\varepsilon_2$ can be easily chosen with a small value of 0.05~0.1, because it is only an assistant modification and main problems are solved by methods discussed in above sections.

Considering the rotated Riemann solver for $\xi$ the Eq. (67) can be improved as follows:

$$\xi = \max\left(\left|\frac{U_R + U_L}{2}\right| + \frac{\Delta U}{2}, \varepsilon_1 \left[1 - f^8(\bar{M})\right] f^8(M) f_{rr}\right), \tag{76}$$

$$f_{rr} = |\alpha_1 U_1| + |\alpha_2 U_2|, \tag{77}$$

$$U_1 = n_{1x} u + n_{1y} v + n_{1z} w, \tag{78}$$

$$U_2 = n_{2x} u + n_{2y} v + n_{2z} w, \tag{79}$$

$$\alpha_1 = \boldsymbol{n}_1 \cdot \boldsymbol{n} = n_{1x} n_x + n_{1y} n_y + n_{1z} n_z, \tag{80}$$

$$\alpha_2 = \boldsymbol{n}_2 \cdot \boldsymbol{n} = n_{2x} n_x + n_{2y} n_y + n_{2z} n_z, \tag{81}$$

where $\boldsymbol{n}$ is normal direction of the cell face, and the rotated direction $\boldsymbol{n}_1$ can be determined according to Ref. [23]:

$$\boldsymbol{n}_1 = \begin{cases} \boldsymbol{n} & \text{if } \sqrt{(\Delta u)^2 + (\Delta v)^2 + (\Delta w)^2} < 10^{-5} U_{ref}, \\ \dfrac{\Delta u \boldsymbol{i} + \Delta v \boldsymbol{j} + \Delta w \boldsymbol{k}}{\sqrt{(\Delta u)^2 + (\Delta v)^2 + (\Delta w)^2}} & \text{otherwise,} \end{cases} \tag{82}$$

and $n_2$ is perpendicular to $n_1$:

$$n_2 = (n_1 \times n) \times n_1. \tag{83}$$

The disadvantage of the modified entropy fix in Eq. (74) is set a minimum value for each cell face, and it can be avoided by the rotated Riemann solver in Eq. (76). However, the rotated Riemann solver may produce unnecessary large value of $f_{rr}$ near shock, which can be limited by the entropy fix. Therefore, the modified entropy fix and the rotated Riemann solver can be combined with complementary advantages as follows:

$$\xi = \max\left\{\left|\frac{U_R + U_L}{2}\right| + \frac{\Delta U}{2}, \varepsilon_1\left[1 - f^8(\bar{M})\right]f^8(M)\min\left(f_{ef}, f_{rr}\right)\right\}. \tag{84}$$

When $\varepsilon_1 = 0$, Eq. (84) becomes Eq. (67) for general flows. If necessary for difficult computation, the value of $\varepsilon_1$ can be set to 1 for more robust by increasing limited numerical dissipation.

## 6. An Improved Roe Scheme for All Mach-Number Flows Simultaneously Curing Known Problems

### 6.1 Some Remarks for Problems of Simulating All Mach-Number Flows

In Eq. (7), the numerical dissipation is subdivided into five terms. Besides of the basic upwind dissipation term $\xi$, the terms $\delta p_p$, $\delta p_u$, $\delta U_u$ and $\delta U_p$ are velocity-difference-driven and pressure-difference-driven direct and cross modifications for the cell face pressure and velocity, respectively, as shown in Fig.2.

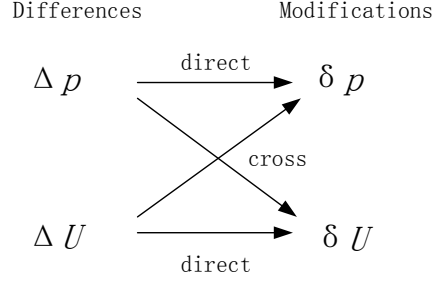

Fig.2 The relation between differences and modifications

It is not surprising that direct modifications $\delta p_p$ and $\delta U_u$ are not easy to cause problems, but not vice verse. In fact, the cross modifications $\delta p_u$ and $\delta U_p$ play very important role in the failings of the Roe scheme for all Mach-number flows as discussed in the above paragraphs.

Further, because of quasi one-dimensional characteristic, the multi-dimensional Roe scheme is constructed based on normal velocity $U$ or normal Mach number $\bar{M}$ on the cell face. For the one-dimensional calculation, $\bar{M}$ is equal to Mach number $M$. It is reason why many fails, such as non-physical behavior problem for incompressible flows and shock instability problem for compressible flows, do not occur for one-dimensional simulation.

For the multi-dimensional calculation, however, $\bar{M}$ is not equal to $\bar{M}$. In detail, there exist following two different conditions:

(1) $\bar{M} \to 0$ for incompressible flows, i.e., $M \to 0$;

(2) $\bar{M} \to 0$ for compressible flows, i.e., $M \geq 1$ or pressure-density fluctuation is large;

The two conditions have different physical behaviour and simulation requirement, especially for the cross modifications. For example, the former requires that $\delta p_u \leq \mathrm{O}(\rho u \Delta u)$ and $\delta U_p = \mathrm{O}(c^{-1}\rho^{-1}\Delta p \sim u^{-1}\rho^{-1}\Delta p)$, the latter requires that $\delta p_u = \mathrm{O}(\rho c \Delta u)$ and $\delta U_p = 0$. However, the classical Roe scheme confuses the two conditions because it has only a criterion of $\bar{M}$. Thus, for multi-dimensional flows non-physical behavior, checkerboard and shock instability problems occur.

Therefore, to a great extent known problems of simulating all Mach-number flows are due to one reason: when $\bar{M} \to 0$ incompressible and compressible flows cannot be identified, which lead to incorrect cross modifications.

This explanation also provides possibility of simultaneously curing known problems with simple corrections. As long as introducing Mach number and an assistant pressure-density-varying detector if necessary into the scheme, compressible and incompressible flows can be identified and thus cross modifications can be ensured correct.

Besides of these four terms, the basic upwind dissipation term $\xi$ is also discovered as the reason of positive non-conservativeness, which is also regarded as an important reason causing shock instability as shown in numerical cases in this paper. Additional, for more robust if necessary, the term $\xi$ can be limitedly increased by combing the modified entropy fix and the rotated Riemann solver with complementary advantages.

**6.2 An Improved Roe Scheme for All Mach-Number Flows Simultaneously Curing Known Problems**

Summarizing above sections, the Roe-AM scheme can be proposed for all Mach-number flows simultaneously curing almost all known problems. Considering other scalar equations, such as the turbulence model, the complete form of the Roe-AM scheme is expressed as follows:

$$\tilde{F}_d^{Roe-AM-TP} = -\frac{1}{2}\left\{\xi\begin{bmatrix}\Delta\rho\\\Delta(\rho u)\\\Delta(\rho v)\\\Delta(\rho w)\\\Delta(\rho H)\\\Delta(\rho k)\\\vdots\end{bmatrix} + (\delta p_p + \delta p_u)\begin{bmatrix}0\\n_x\\n_y\\n_z\\0\\0\\\vdots\end{bmatrix} + \delta U_p\begin{bmatrix}\rho\\\rho u\\\rho v\\\rho w\\\rho H\\\rho k\\\vdots\end{bmatrix} + \delta U_{u,R}\begin{bmatrix}\rho\\\rho u\\\rho v\\\rho w\\\rho H\\\rho k\\\vdots\end{bmatrix}_R + \delta U_{u,L}\begin{bmatrix}\rho\\\rho u\\\rho v\\\rho w\\\rho H\\\rho k\\\vdots\end{bmatrix}_L\right\}$$

, (85)

$$\xi = \max\left\{\left|\frac{U_R+U_L}{2}\right| + \frac{\Delta U}{2}, \varepsilon_1\left[1-f^8(\bar{M})\right]f^8(M)\min(f_{ef}, f_{rr})\right\}, \quad (86)$$

$$\delta p_p = \text{sign}(U)\min(|U|', c)\frac{\Delta p}{c}, \quad (87)$$

$$\delta p_u = \left[1 - s_1 + s_1 f(M)\right]\max(0, c - |U|')\rho\Delta U, \quad (88)$$

$$\delta U_p = s_1\left[1 - f^8(M)\right]\left[\max(0, \tilde{c} - |U|') + (1-\theta)\left(|U|' - \frac{U}{2\tilde{c}}\text{sign}(\tilde{U})\min(|\tilde{U}|, \tilde{c})\right)\right]\frac{\Delta p}{\rho\theta c^2}$$

, (89)

$$\delta U_{u,R} = \frac{1}{2}\text{sign}(U)\min(|U|', c)\frac{\Delta U}{c}, \quad (90)$$

$$\delta U_{u,L} = \frac{1}{2}\text{sign}(U)\min(|U|', c)\frac{\Delta U}{c}. \quad (91)$$

$$|U|' = |U| - f(\bar{M})\frac{\text{sign}(U+c)\max(0, U_R - U_L) - \text{sign}(U-c)\max(0, U_R - U_L)}{4}, (92)$$

$$M = 0.5\left(\frac{\sqrt{u_L^2 + v_L^2 + w_L^2}}{c_L} + \frac{\sqrt{u_R^2 + v_R^2 + w_R^2}}{c_R}\right), \quad (93)$$

$$\bar{M} = \frac{|U|}{c}, \quad (94)$$

$$s_1 = f^8(a), \quad (95)$$

$$a_{i+1/2,j,k} = \min\left(\begin{array}{c} P_{i+1/2,j,k}, P_{i+1,j-1/2,k}, P_{i+1,j+1/2,k}, P_{i,j-1/2,k}, P_{i,j+1/2,k}, \\ P_{i+1,j,k-1/2}, P_{i+1,j,k+1/2}, P_{i,j,k-1/2}, P_{i,j,k+1/2} \end{array}\right), \tag{96}$$

$$P_{i+1/2,j,k} = \min\left(\frac{p_{i,j,k}}{p_{i+1,j,k}}, \frac{p_{i+1,j,k}}{p_{i,j,k}}, \frac{\rho_{i,j,k}}{\rho_{i+1,j,k}}, \frac{\rho_{i+1,j,k}}{\rho_{i,j,k}}\right), \tag{97}$$

$$f(\phi) = \min\left(\phi \frac{\sqrt{4+(1-\phi^2)^2}}{1+\phi^2}, 1\right), \tag{98}$$

$$\tilde{c} = \frac{1}{2}\sqrt{4c^2\theta + (1-\theta)^2 U^2}, \tag{99}$$

$$\tilde{U} = \frac{1}{2}(1+\theta)U, \tag{100}$$

$$\theta = \min\left[\max\left(M_{ref}^2, M^2\right), 1\right]. \tag{101}$$

$$f_{ef} = \varepsilon_2 c, \tag{102}$$

$$f_{rr} = |\alpha_1 U_1| + |\alpha_2 U_2|, \tag{103}$$

$$U_1 = n_{1x}u + n_{1y}v + n_{1z}w, \tag{104}$$

$$U_2 = n_{2x}u + n_{2y}v + n_{2z}w, \tag{105}$$

$$\alpha_1 = \boldsymbol{n}_1 \cdot \boldsymbol{n} = n_{1x}n_x + n_{1y}n_y + n_{1z}n_z, \tag{106}$$

$$\alpha_2 = \boldsymbol{n}_2 \cdot \boldsymbol{n} = n_{2x}n_x + n_{2y}n_y + n_{2z}n_z, \tag{107}$$

$$\boldsymbol{n}_1 = \begin{cases} \boldsymbol{n} & \text{if } \sqrt{(\Delta u)^2 + (\Delta v)^2 + (\Delta w)^2} < 10^{-5} U_{ref}, \\ \dfrac{\Delta u \boldsymbol{i} + \Delta v \boldsymbol{j} + \Delta w \boldsymbol{k}}{\sqrt{(\Delta u)^2 + (\Delta v)^2 + (\Delta w)^2}} & \text{otherwise,} \end{cases} \tag{108}$$

$$\boldsymbol{n}_2 = (\boldsymbol{n}_1 \times \boldsymbol{n}) \times \boldsymbol{n}_1. \tag{109}$$

where $\phi$ in Eq. (98) represents any scalar variable.

The classical Roe scheme and the Roe-AM scheme are compared as follows. For satisfying full upwind and positivity conditions when $U \geq c$, the term $\delta U_u$ is subdivided into $\delta U_{u,R}$ in Eq. (90) and $\delta U_{u,L}$ in Eq. (91), and term $\xi$ is modified, as shown in the first term on the right side of Eq. (86). To satisfy the positivity condition when $U \to 0$, the term $\xi$ is further improved as shown in Eq. (86), and term $\delta U_p$ is improved while considering the requirement of suppressing the checkerboard problem for incompressible flows, as shown in Eq. (89). If necessary, the term $\xi$ also provides a limited dissipation combing the modified entropy fix and the rotated Riemann solver in Eq. (86). To address expansion shock, $|U|$ in Eqs. (87)–(91) is replaced by $|U|'$ defined in Eq. (92). To resolve the non-physical behavior problem for incompressible flows, the term $\delta p_u$ is improved as shown in Eq. (95).

In the Roe-AM scheme, there exist three tunable values: a switch function $s_1$, a switch parameter $\varepsilon_1$, and a empirical parameter $\varepsilon_2$.

The function $s_1$ in Eq. (95) considers the possibility of low-Mach-number large-pressure-gradient flows. Without this possibility, the coefficients can be ignored to save computational cost by setting the following variable:

$$s_1 = 1. \tag{110}$$

The switch parameter $\varepsilon_1$ can be set as the following variable:

$$\varepsilon_1 = \begin{cases} 0 & \text{for general conditions} \\ 1 & \text{for better robust} \end{cases}. \tag{111}$$

Because preceding improvements can solve most of problems, $\varepsilon_1$ can be zero for general conditions to save computational cost. If necessary for difficult simulations, $\varepsilon_1$ can be one to introduce modification combing the entropy fix and the rotated Riemann solver for better robust.

The parameter $\varepsilon_2$ is empirical similar to that of the entropy fix, but it can be easily chosen because it only play an assistant role. The value of $\varepsilon_2$ is suggested as follows:

$$\varepsilon_2 = 0.05. \tag{112}$$

Therefore, through subtle improvements based on the classical Roe scheme, the Roe-AM scheme can simultaneously overcome almost all well-known drawbacks of the classical Roe scheme with minimal numerical dissipation increase.

## 7. Numerical Tests

### 7.1 Shock Tube Tests

For shock tube simulation, mesh grids are set as $200 \times 10$ along the streamwise and normal directions. The four-stage Runge–Kutta scheme is adopted for time discretization, and first-order accuracy is adopted for space discretization to discuss the schemes themselves. For Roe-AM scheme, $\varepsilon_1 = 0$ unless otherwise specified.

#### 7.1.1 Expansion Shock Test

Initial conditions of the shock tube are expressed as $\rho_L = 3$, $w_L = 0.9$, $p_L = 3$, $\rho_R = 1$, $w_R = 0.9$, and $p_R = 1$ at the streamwise position of $z = 0.3$.

As shown in Fig. 3(a), the Roe scheme evidently produces an non-physical expansion shock at the $z = 0.3$ position and at $t = 0.2$ s. As shown in Figs. 3(b) and (c), the Roe-AM scheme without improvement in Eq. (92) cannot resolve expansion shock, but the complete version of the Roe-AM scheme can. This condition validates the effectiveness of suppressing the expansion shock by redefining $|U|$ in Eq. (92).

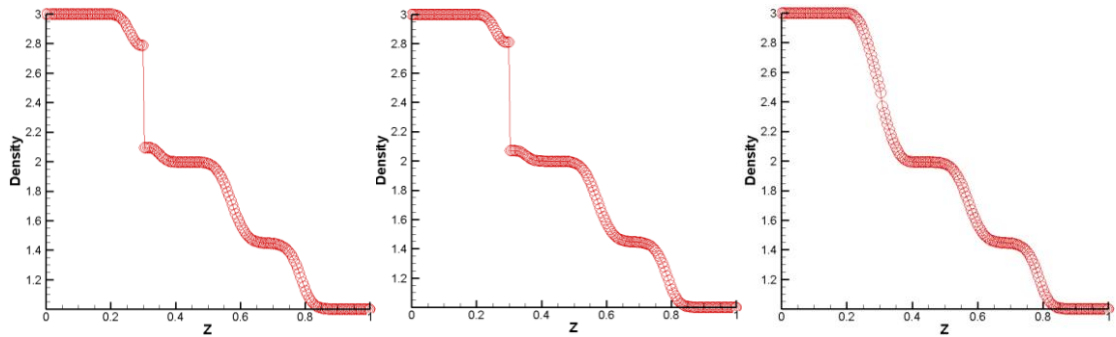

(a) Roe      (b) Roe-AM without improvement in Eq. (92)   (c) Roe-AM

Fig. 3 Results of the test at $t = 0.2\,\mathrm{s}$

### 7.1.2 Near-vacuum Test

Initial conditions of the test are as follows: $\rho_L = 1$, $w_L = -2$, $p_L = 0.4$, $\rho_R = 1$, $w_R = 2$, and $p_R = 0.4$ at the streamwise position of $z = 0.5$. This test produces a very small pressure and a density area close to vacuum between two strong rarefactions. The near-vacuum area leads to computation difficulties for schemes. The Roe scheme fails to converge for this test. The Roe-AM scheme can obtain reasonable results because of the satisfied positivity condition, as shown in Fig. 4.

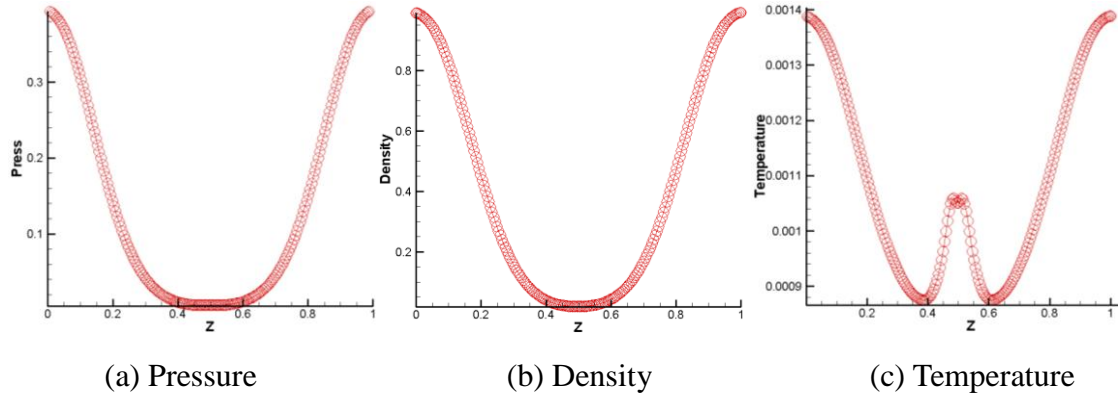

(a) Pressure      (b) Density      (c) Temperature

Fig. 4 Results of the test at $t = 0.15\,\mathrm{s}$ by the Roe-AM scheme

### 7.1.3 Strong Shock Test

Initial conditions of the test are as follows: $\rho_L = 1$, $w_L = 0$, $p_L = 1000$, $\rho_R = 1$, $w_R = 0$, and $p_R = 0.01$ at the streamwise position of $z = 0.5$. It is actually the left half of the blast wave problem [33] and is a very severe case for testing robustness of a scheme. As shown in Fig. 5, the Roe-AM scheme features the same performance as the Roe scheme. This result indicates that modifications in the Roe-AM scheme exert no effect on its robustness.

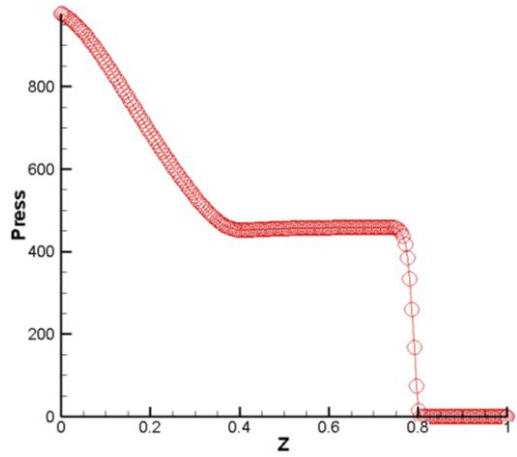
(a) Pressure distribution by Roe

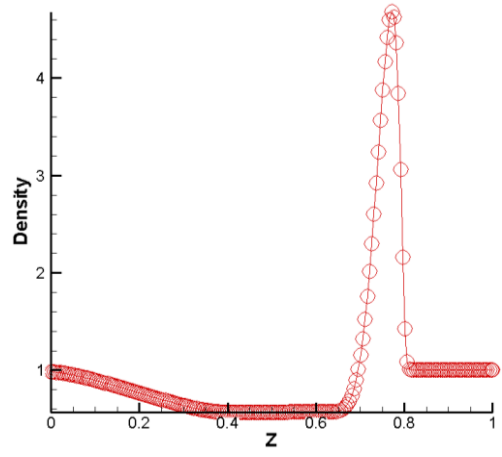
(b) Density distribution by Roe

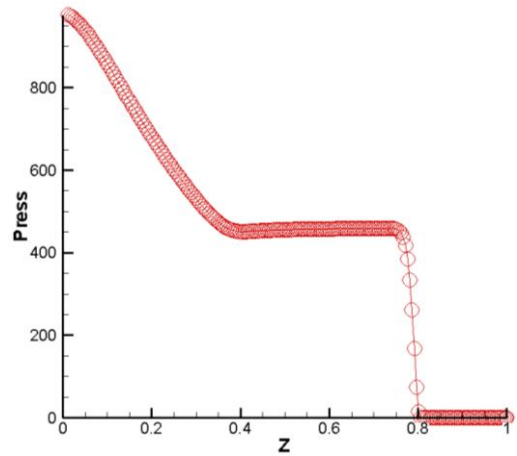
(c) Pressure distribution by Roe-AM

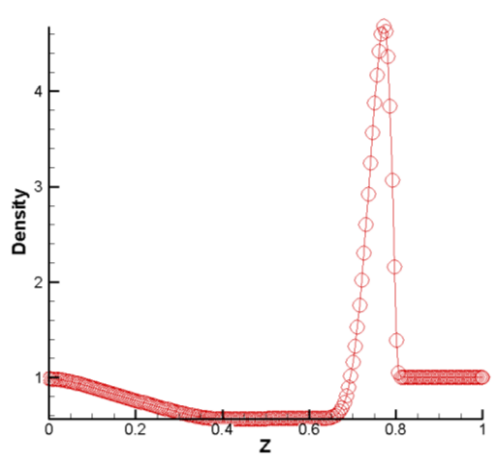
(d) Density distribution by Roe-AM

Fig. 5 Results of the test at $t = 0.012$ s

### 7.2 Odd–Even Decoupling Test

Odd–even decoupling test is an important case designed by Quirk [19], because any scheme that suffers it also suffers from shock instability in other classical cases. The initial conditions are given as $(\rho, p, u, v)_L = \left( \frac{1512}{205}, \frac{251}{6}, \frac{175}{36}, 0 \right)$ and $(\rho, p, u, v)_R = (1.4, 1, 0, 0)$. Therefore, a planar shock moves with the Mach number of 6 in a duct. The computational mesh includes $20 \times 800$ right orthogonal uniform grids in the $Y$ and $X$ directions, except that the centerline grid is odd–even disturbed as follows:

$$Y_{i,j,\text{mid}} = \begin{cases} Y_{j,\text{mid}} + \varepsilon_y \Delta Y, & \text{for } i \text{ even,} \\ Y_{j,\text{mid}} - \varepsilon_y \Delta Y, & \text{for } i \text{ odd.} \end{cases} \quad (113)$$

In this test, a large value of $\varepsilon_y = 0.1$ is adopted because it produces a more serious odd–even decoupling than small values.

Apart from the original Roe scheme and its improvement in this paper, the following entropy fix in Eqs. (114) and (115) are also considered for comparison because it is commonly used.

$$\lambda_i = \begin{cases} \lambda_i, & \lambda_i \geq h, \\ \dfrac{1}{2}\left(\dfrac{\lambda_i^2}{h} + h\right), & \lambda_i < h, \end{cases} \quad (114)$$

$$h = \varepsilon_\lambda \max(\lambda_i), \quad (115)$$

where $\varepsilon_\lambda$ is a constant with a commonly adopted value of 0.05 to 0.2.

As shown in Fig. 6(a), moving shock is smoothed and destroyed at 100 s by the Roe scheme. Adopting the entropy fix, the result only achieves a minor improvement with $\varepsilon_\lambda = 0.05$, and shock is also seriously deformed even when $\varepsilon_\lambda = 0.2$, as shown in Fig. 6(b) and (c).

Adopting the Roe-AM scheme without considering improvements for the pressure-positivity term in Eq. (89) and velocity-positivity term in Eq. (86), that is, $\xi$ and $\delta U_p$ terms adopt the definitions of the Roe scheme, and other terms adopt the definitions of the Roe-AM scheme, producing the results in Fig. 6(d), and these findings are very similar to those in Fig. 6(a) which are obtained through the Roe scheme. These results also indicate that the improvements, except the positivity, cause little effect on shock instability problem. As shown in Figs. 6(e) and 6(f), pressure- and velocity-positivity modifications can remarkably improve shock instability, and pressure positivity is more important than velocity positivity.

Adopting the Roe-AM scheme, shock instability is nearly resolved, but a small oscillation remains, as shown in Fig. 6(g). To discuss the cause of small oscillation, the AUSM+ scheme, which is regarded as the best scheme for suppressing shock instability, is adopted to obtain the results shown in Fig. 6(h). Compared with Figs. 6(g) and 6(h), two different schemes produce surprisingly similar results with small oscillation. Therefore, there still exists unclear reason that causes weak shock instability.

Although it is a pity that not all mechanisms of shock instability are discovered, the remaining weak shock instability can be cured by adding limited numerical dissipation into the term $\xi$. When $\varepsilon_1 = 1$ and $\varepsilon_2 = 0.05$, the Roe-AM scheme obtain good result as shown in Fig. 6(i).

Fig. 7 shows contours of rotated Mach number $f_{rr}/c$ on cell faces parallel or nearly parallel to $X$ direction based on results in Fig. 6(i). It is clear that rotated Mach number is nearly zero on smooth area, which is better than the entropy fix. On odd–even disturbed centerline grids and some areas near shock, however, values are large and maximum value is up to 1.73. Such values seem too large although they make the scheme very robust. It is the reason why the entropy fix and rotated Riemann solver have good complementary advantages.

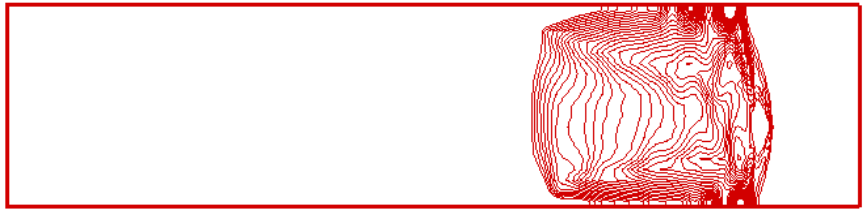

(a) Roe

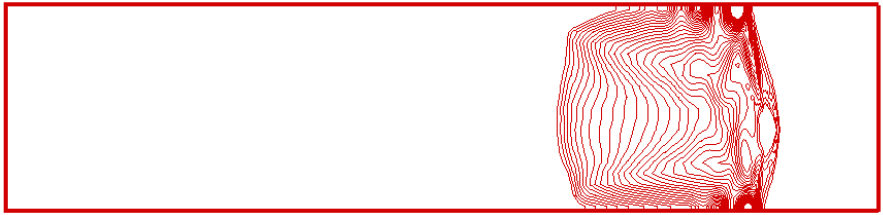

(b) Roe with the entropy fix $\varepsilon_\lambda = 0.05$

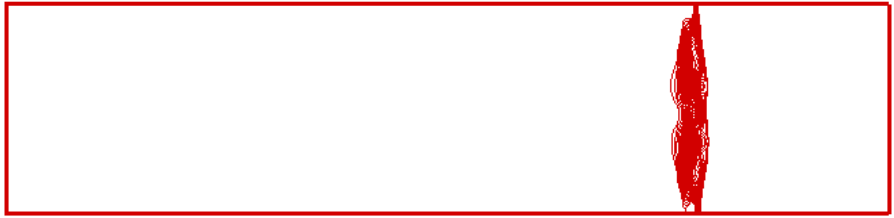

(c) Roe with the entropy fix $\varepsilon_\lambda = 0.2$

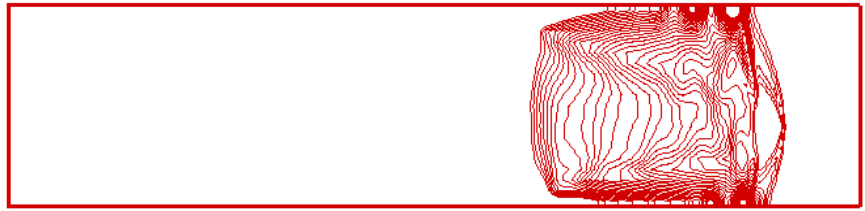

(d) Roe-AM without the modification for pressure- and velocity-positivity terms

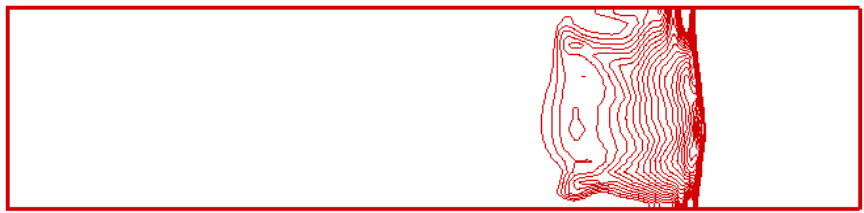

(e) Roe-AM without the modification for pressure-positivity term

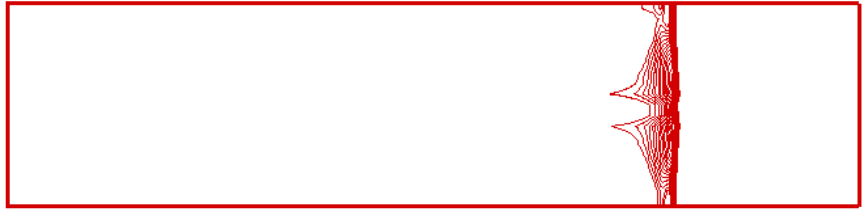

(f) Roe-AM without the modification for velocity-positivity term

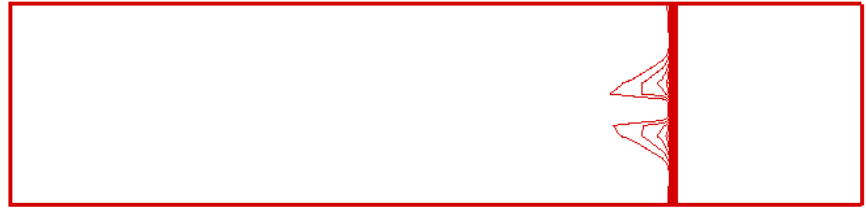

(g) Roe-AM

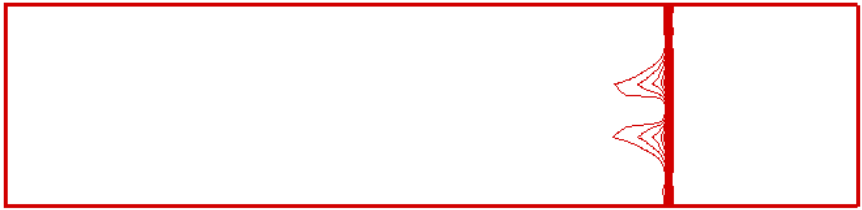

(h) AUSM+

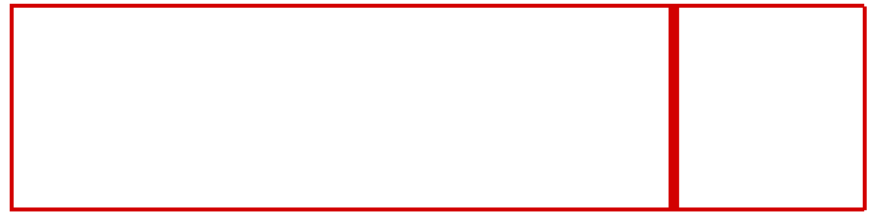

(i) Roe-AM with $\varepsilon_1 = 1$ and $\varepsilon_2 = 0.05$

Fig. 6 Density contours of the odd–even decoupling test at $t = 100$ s

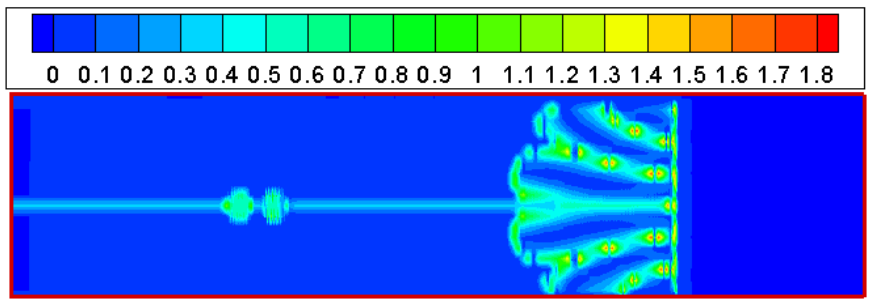

Fig. 7 Contours of rotated Mach number $f_{rr}/c$ on cells nearly parallel to flows

## 7.3 Supersonic Corner Test

The supersonic corner test considers a moving supersonic shock around a 90° corner. In addition to shock instability, an expansion shock maybe produced through numerical computation around the corner where a series of expansion waves exist.

Initial conditions of the supersonic corner test are as follows: $\rho_L = 7.04108$, $u_L = 4.07794$, $v_L = 0$, $p_L = 30.05942$, $\rho_R = 1.4$, $u_R = v_R = 0$, and $p_R = 1$ at the $x$-axis position of 0.05. With $400 \times 400$ grids, the Roe scheme produces instability at the top right corner of the shock, as shown in Fig. 8(a). Adopting the entropy fix with large value of $\varepsilon_\lambda = 0.2$, the instability is improved but still obvious, as shown in Fig. 8(b).

Adopting Roe-AM without the modification for velocity-positivity term in Eq. (86), the result in Fig. 8(c) is similar to that in Fig. 8(b).

Adopting the Roe-AM scheme without the improvement for expansion shock in Eq. (92), shock instability is observably fixed, which demonstrate that velocity-positivity condition is play an important role in shock instability again. However, visible expansion shock occurs, as shown in Fig. 8(d). This result indicates that non-physical shock is also held when physical shock is held.

Adopting the complete version of the Roe-AM scheme, shock instability and expansion shock are fixed simultaneously as expected. However, insignificant shock obscure can be observed as shown in Fig. 8(d) and (e), which indicates that there still exists unclear mechanism as discussed for Fig. 6(g). Adopting $\varepsilon_1 = 1$ and $\varepsilon_2 = 0.05$, this insignificant instability can be cured as shown in Fig. 8(f).

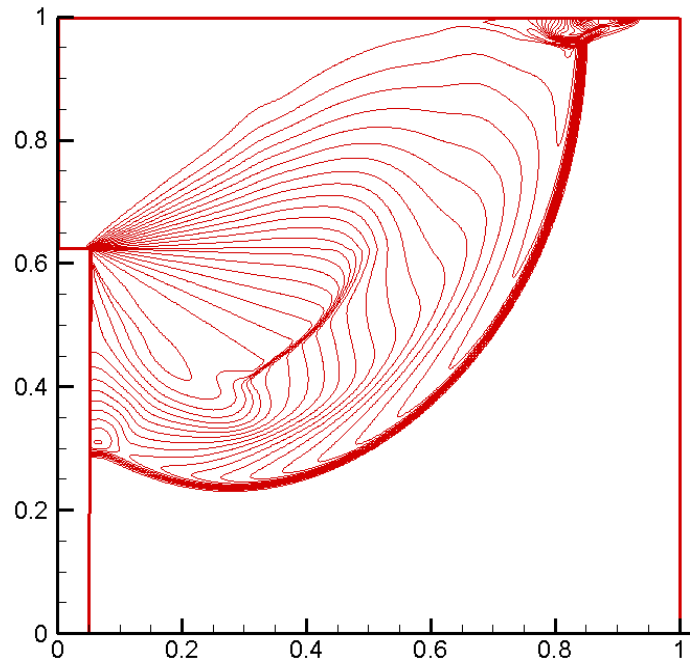

(a) Roe

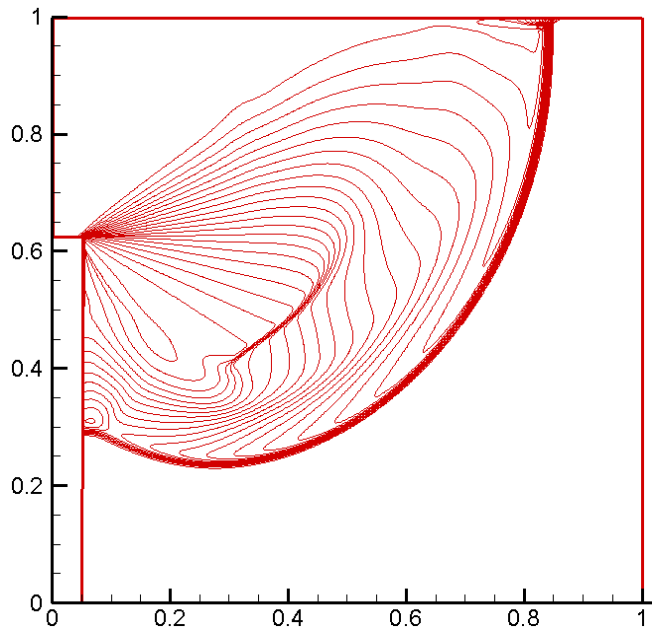

(b) Roe with the entropy fix $\varepsilon_\lambda = 0.2$

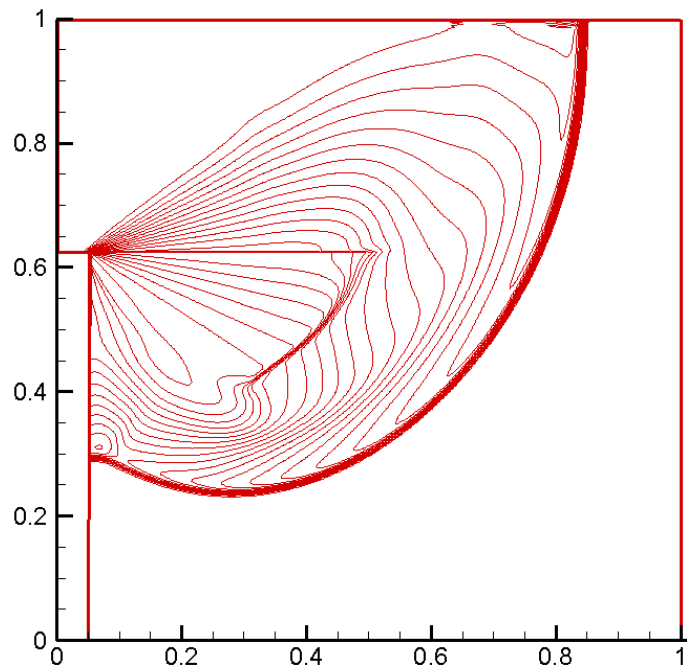

(c) Roe-AM without the modification for velocity-positivity term

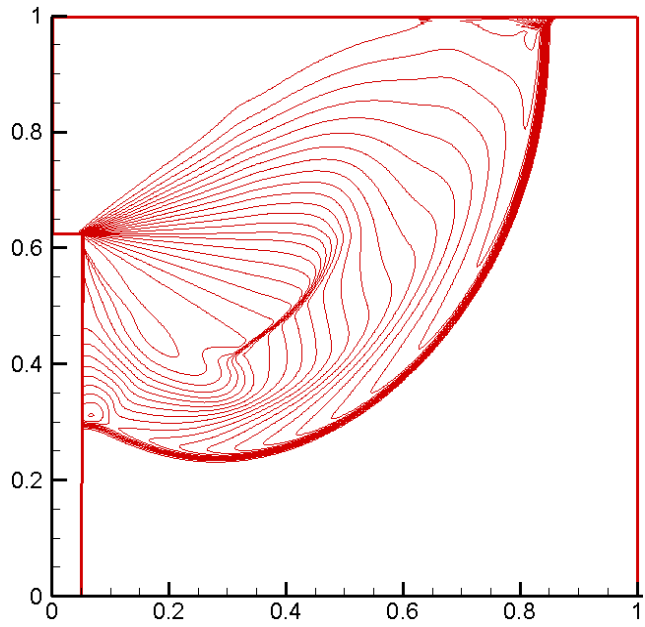

(d) Roe-AM without expansion-shock improvement in Eq. (92)

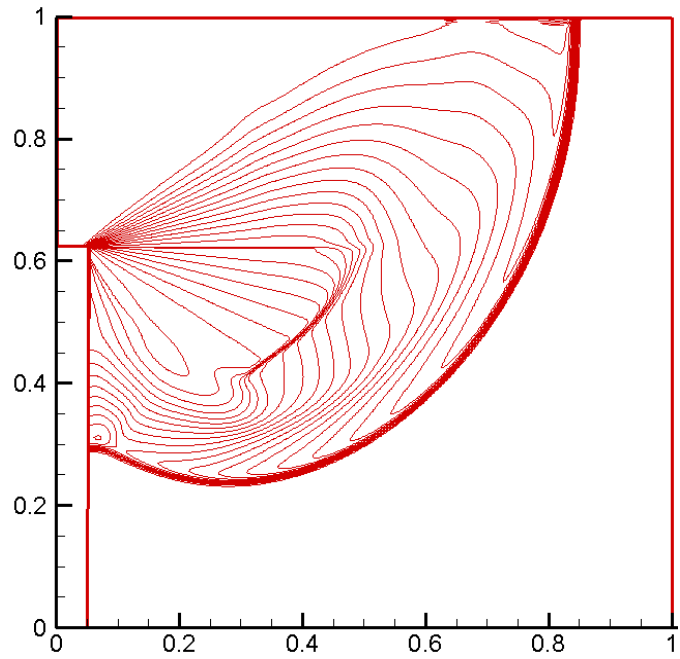

(e) Roe-AM

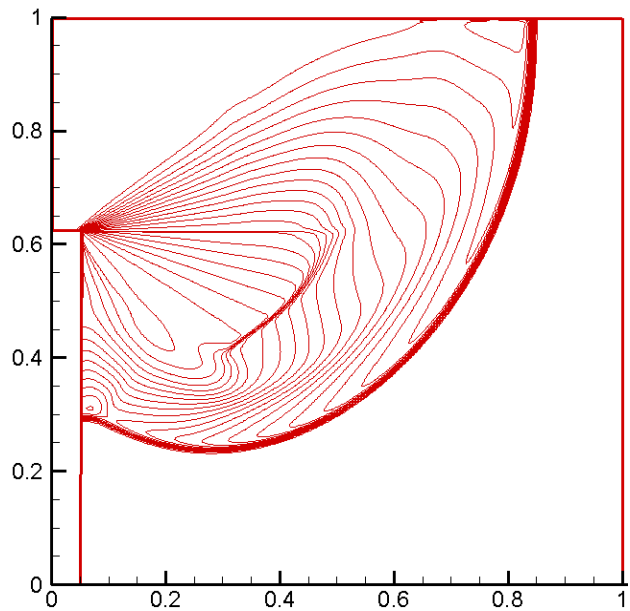

(f) Roe-AM with $\varepsilon_1 = 1$ and $\varepsilon_2 = 0.05$

Fig. 8 Density contours of the supersonic corner test at $t = 0.155\,\text{s}$

## 7.4 Kinked Mach Stem of the Double-Mach Reflection Test

The kinked Mach stem of the double-Mach reflection test is another well-known shock instability problem. This problem occurs when an inclined moving shock is reflected from a wall. Shock is initially set up at an inclination angle of 60 ° with a Mach number of 10.

In Fig. 9, density contours are shown at 0.2 s on 200 × 800 grids. For the Roe scheme, shock is severely deformed, and a non-physical triple point appears; these conditions make up the kinked Mach stem, as shown in Fig. 9(a). For the Roe-AM scheme, the kinked Mach stem is fixed as expected in Fig. 9(b). This case is relatively easy, because the shock instability can be cured as Fig. 9(b) even $\varepsilon_1 = 0$ and velocity-positivity condition is unsatisfied.

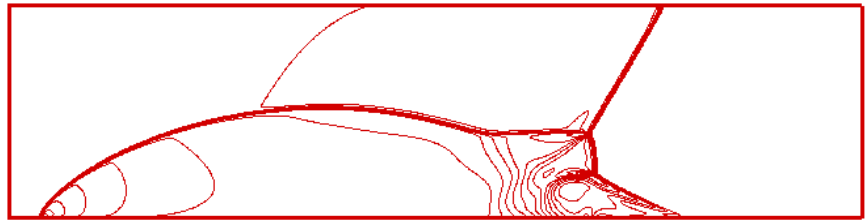

(a) Roe

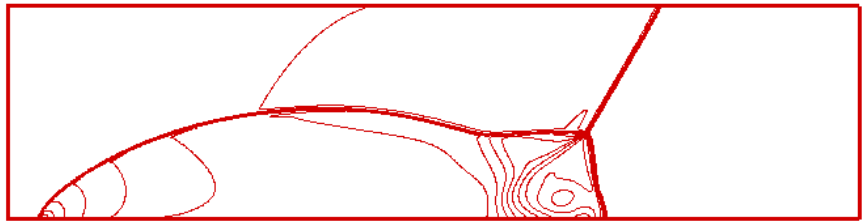

(b) Roe-AM

Fig. 9 Density contours of the double-Mach reflection test at $t = 0.2\,\text{s}$

## 7.5 Carbuncle of the Supersonic Flow Test around a Circular Cylinder

The carbuncle phenomenon is also a well-known form of shock instability for supersonic flows around a circular cylinder with a free-stream Mach number of 20. With 20 × 160 grids in the radial and circumferential directions, the carbuncle phenomenon addressed by the Roe scheme is shown in Fig. 10(a), and that fixed by the Roe-AM scheme is shown in Fig. 10(b). It is also a relatively easy case, and it can obtain good results as Fig. 10(b) even to set $\varepsilon_1 = 0$, $s_1 = 1$ and velocity-positivity condition unsatisfied.

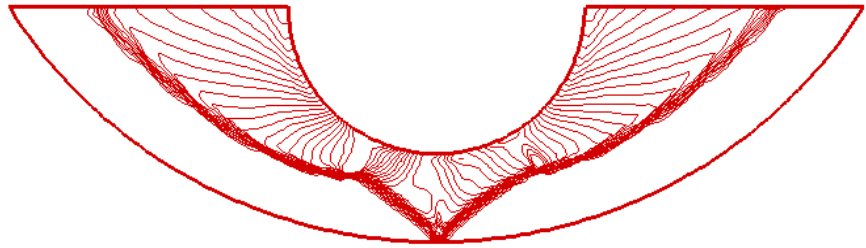

(a) Roe

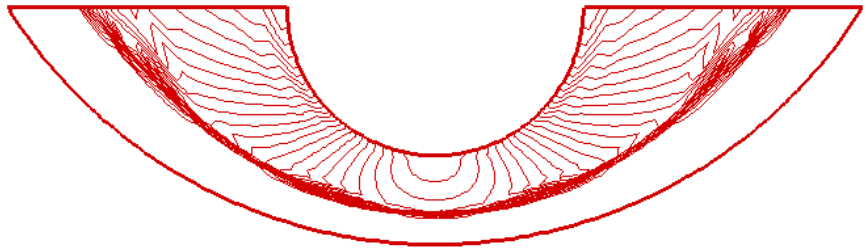

(b) Roe-AM

Fig. 10 Pressure contours of the carbuncle test

**7.6 Low-Mach-Number Inviscid Flow Test around a Cylinder**

The two-dimensional inviscid flow that passes a cylinder is a typical test case for low Mach number. In this test inflow, O-type grids are $100 \times 72$ along the circumference and radius.

For flows with $M = 0.01$, as shown in Fig. 11(a), the Roe scheme produces a non-physical solution that resembles full-viscous Stokes flow for inviscid flows. Adopting the Roe-AM scheme with $\delta U_p = 0$, the solution suffers from a serious checkerboard problem, as shown in Fig. 11(b). This serious checkerboard rapidly appears even if the initial solution is good, and increases leading to computational divergence.

Adopting the Roe-AM scheme with $\delta U_p = \max(0, c - |U|)\frac{\Delta p}{\rho c^2}$, which is the term of Roe scheme in Eq. (17), some weak checkerboard can be observed in Fig. 11(c). This phenomenon accords with theoretical analysis in section 3.2.

Adopting the complete version of the Roe-AM scheme, the correct solution is obtained and all checkerboard modes are suppressed, as shown in Fig. 11(d). When the Mach number decreases to 0.001, the Roe-AM scheme can also produce good result, and non-dimensional pressure contours in Fig. 11(e) are the same as that in Fig. 11(d).

Following Ref. [4], Fig. 12 shows pressure fluctuations $\text{Ind}(p) = (P_{max} - P_{min}) / P_{max}$ versus the inlet Mach number. This result agrees perfectly with theoretical asymptotic predictions: pressure fluctuations scale exactly with $M_*^2$, which can be recovered by the Roe-AM scheme, but the Roe scheme produces non-physical pressure fluctuations that scale with $M_*$.

Therefore, this test validates the improvement for $\delta p_u$ term in Eq. (88) for suppressing the non-physical behavior problem and the necessity of term $\delta U_p$ in Eq. (89) for suppressing the checkerboard problem.

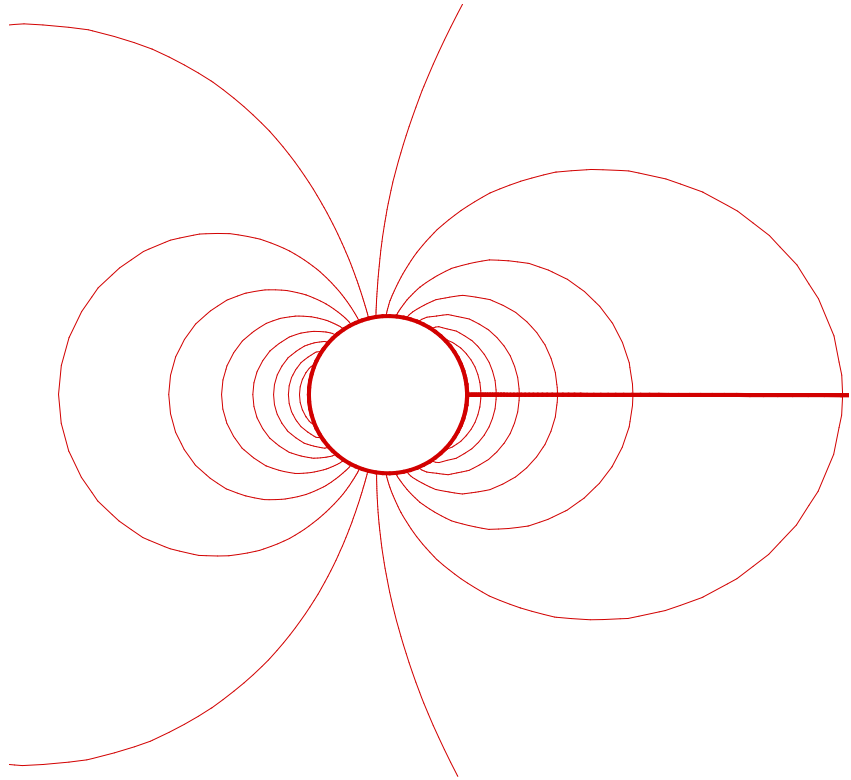

(a) Roe for flows of $M = 0.01$

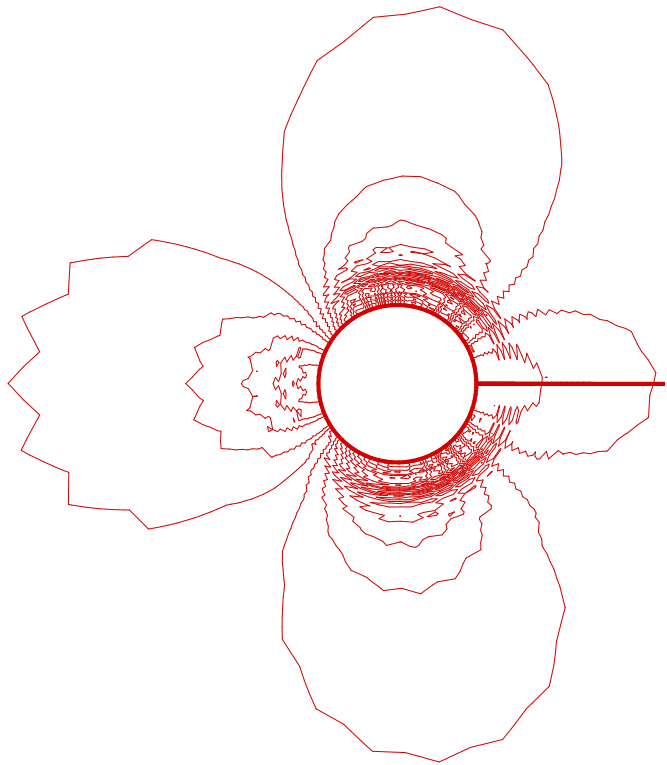

(b) Roe-AM with $\delta U_p = 0$ for flows of $M = 0.01$

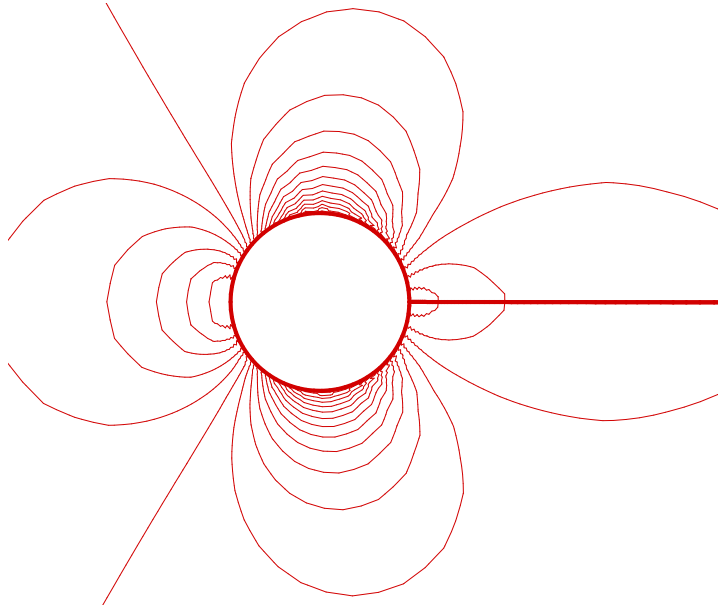

(c) Roe-AM with $\delta U_p = \max\left(0, c-|U|\right)\dfrac{\Delta p}{\rho c^2}$ for flows of $M = 0.01$

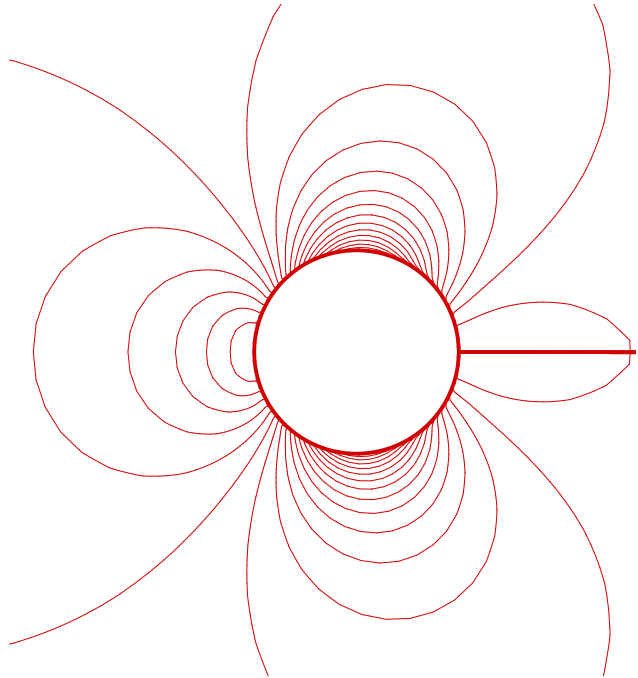

(d) Roe-AM for flows of $M = 0.01$

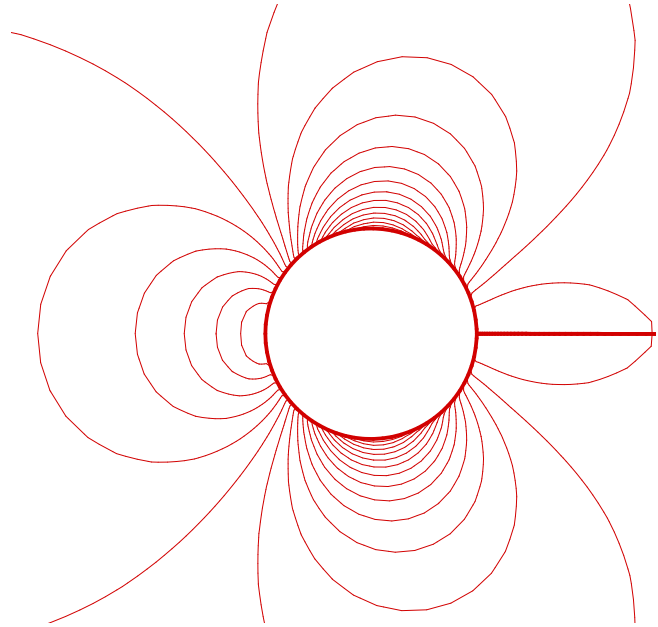

(e) Roe-AM for flows of $M = 0.001$

Fig. 11 Pressure contours of low-Mach-number inviscid flows around a cylinder

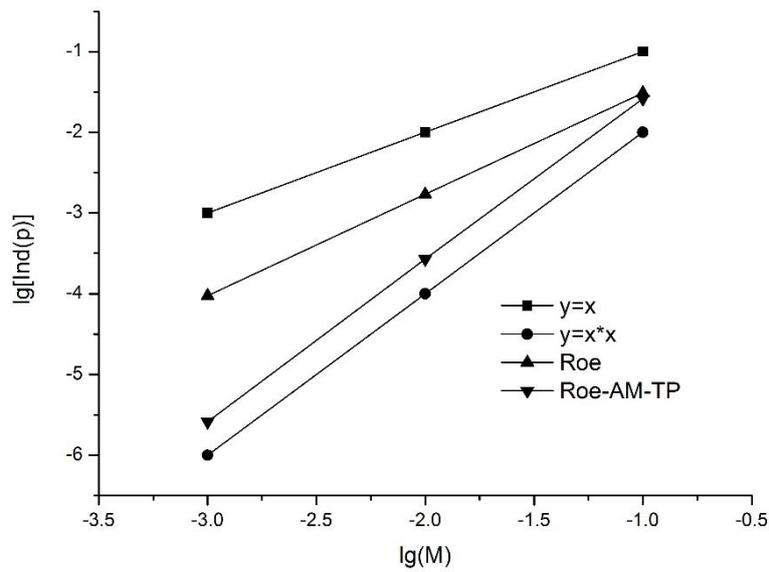

Fig. 12 Pressure fluctuations vs inflow Mach number

## 8. Conclusions

In this paper, drawbacks of the Roe scheme mechanisms are analyzed for extremely low- and high-Mach-number flows, and corresponding improvements are discussed and proposed, which are summed up as follows:

(1) For all Mach-number flows, main problems including non-physical behavior, checkerboard, and most of shock instability are due to one reason: incompressible and compressible flows is confused and thus incorrect cross modifications are produced when normal Mach number on the cell face tends to zero. This discovery makes it possible to simultaneously cure known problems by a simple method, which introduces Mach number and an assistant pressure-density-varying detector into the Roe scheme to judge compressibility;

(2) The mechanism of the preconditioned Roe scheme of suppressing the checkerboard problem can be introduced into the Roe scheme for better effect;

(3) The velocity-positivity condition is also identified as another important reason for shock instability;

(4) The modified entropy fix and the rotated Riemann solver is combined with complementary advantages as an assistant improvement for better robust by modifying the basic upwind dissipation;

(5) Combining previous and current improvements, the Roe-AM scheme is proposed for all Mach-number flows.

According to theoretical analysis and numerical validation, through several subtle adjustments based on the classical Roe scheme, the Roe-AM scheme features significant advantages; it is simple, easy to implement, computationally low-cost, robust, and is essentially free of empirical parameters. The Roe-AM scheme also exhibits good extensibility, and most importantly, it can simultaneously overcome nearly all well-known drawbacks of the classical Roe scheme, with minimal numerical dissipation increases, and even with possible decreases in numerical dissipation. The Roe-AM scheme seems near perfect but still needs some improvements as follows:

(1) A better mechanism is needed compatible with suppressing checkerboard problem and divergence-free constraint of the leading-order velocity;

(2) There is still unclear for the reason causing some weak shock instability. Although this weak shock obscure is not occur always and can be suppressed by increasing limited numerical dissipation if necessary, it is worth discovering the mechanism further and avoid this problem with more reasonable improvement;

(3) For calculation of viscous flows such as large eddy simulation, the numerical dissipation should be investigated further.

## Acknowledgments

This work is supported by Project 51736008 and 51276092 of the National Natural Science Foundation of China.